\documentclass[conference]{IEEEtran}

\ifCLASSINFOpdf
   \usepackage[pdftex]{graphicx}
\else
\fi
\usepackage{amsmath,amssymb,amsfonts}
\usepackage{braket}
\usepackage{xspace}
\usepackage[dvipsnames]{xcolor}
\usepackage{subfig}
\usepackage[nolist]{acronym}

\usepackage{tikz}
\usetikzlibrary{arrows.meta, positioning}
\usepackage[compact]{titlesec}

\usepackage[compact]{titlesec}
\titlespacing*{\section}{0pt}{0.8ex plus 0.3ex minus 0.2ex}{0.6ex plus 0.2ex}
\titlespacing*{\subsection}{0pt}{0.6ex plus 0.2ex minus 0.1ex}{0.4ex plus 0.2ex}
\titlespacing*{\subsubsection}{0pt}{0.5ex plus 0.2ex minus 0.1ex}{0.3ex plus 0.1ex}

\makeatletter
\g@addto@macro\normalsize{%
  \setlength\abovedisplayskip{4pt plus 2pt minus 2pt}
  \setlength\belowdisplayskip{4pt plus 2pt minus 2pt}
  \setlength\abovedisplayshortskip{2pt plus 2pt minus 1pt}
  \setlength\belowdisplayshortskip{2pt plus 2pt minus 1pt}
}
\makeatother

\usepackage{enumitem}
\setlist{nosep,leftmargin=*,topsep=2pt,partopsep=0pt,itemsep=1pt,parsep=1pt}

\usepackage[font=small,skip=3pt]{caption}

\begin{document}
\title{Comparative Analysis of Differential and Collision Entropy for Finite-Regime QKD in Hybrid Quantum Noisy Channels}

\author{\IEEEauthorblockN{Mouli~Chakraborty \IEEEauthorrefmark{6}, Subhash~Chandra \IEEEauthorrefmark{1},  Avishek~Nag \IEEEauthorrefmark{3}, \\
Trung~Q.~Duong \IEEEauthorrefmark{4}, Merouane Debbah \IEEEauthorrefmark{5},  Anshu~Mukherjee \IEEEauthorrefmark{2},}
\IEEEauthorblockA{\IEEEauthorrefmark{6}School of Computer Science and Statistics, Trinity College Dublin, The University of Dublin, College Green, Dublin 2, Ireland\\
\IEEEauthorrefmark{1} School of Natural Sciences,
Trinity College Dublin, The University of Dublin, College Green, Dublin 2, Ireland\\
\IEEEauthorrefmark{3} School of Computer Science,
University College Dublin, Belfield, Dublin 4, Ireland\\
\IEEEauthorrefmark{4} Faculty of Engineering and Applied Science, Memorial University, St. John’s, NL A1C 5S7, Canada, \\
\IEEEauthorrefmark{5} Department of Computer and Information Engineering, Khalifa University, Abu Dhabi, UAE\\
\IEEEauthorrefmark{2} School of Electrical and Electronic Engineering,
University College Dublin, Belfield, Dublin 4, Ireland\\
Email:  moulichakraborty@ieee.org, anshu.mukherjee@ieee.org}
\vspace{-0.5cm}}

\maketitle

\begin{abstract}

In this work, a comparative study between three fundamental entropic measures, differential entropy, quantum Rényi entropy, and quantum collision entropy for a \ac{HQC} was investigated, where \ac{HQN} is characterized by both discrete and \ac{CV} noise components. Using a \ac{GMM} to statistically model the \ac{HQN}, we construct as well as visualize the corresponding pointwise entropic functions in a given 3D probabilistic landscape. When integrated over the relevant state space, these entropic surfaces yield values of the respective global entropy. Through analytical and numerical evaluation, it is demonstrated that the differential entropy approaches the quantum collision entropy under certain mixing conditions, which aligns with the Rényi entropy for order $\alpha=2$. Within the \ac{HQC} framework, the results establish a theoretical and computational equivalence between these measures. This provides a unified perspective on quantifying uncertainty in hybrid quantum communication systems. Extending the analysis to the operational domain of finite-key QKD, we demonstrated that the same $10\%$ approximation threshold corresponds to an order-of-magnitude change in Eve’s success probability and a measurable reduction in the secure key rate.

\end{abstract}

\IEEEpeerreviewmaketitle



\section{Introduction}




Quantum communication enables secure information transfer through protocols such as quantum key distribution (\ac{QKD}) and teleportation \cite{HanzoSatcomm2019}. Real deployments often utilize hybrid quantum–classical systems \cite{mouli2024}, where the two layers interact continuously, creating challenging modeling and optimization problems for settings such as free-space optics and satellite links \cite{PirandolasatQComm2021,Mouli2024Asymp_QKD_SatComm}. A key tool in this context is quantum entropy, which captures uncertainty, information flow, and correlations in these channels. Unlike classical entropy over discrete events, quantum entropies are defined on states that may be in superposition or entangled. The standard choice is the von Neumann entropy $S(\boldsymbol{\rho}) = -\mathrm{Tr}(\boldsymbol{\rho}\log \boldsymbol{\rho})$ \cite{nielsen_chuang_2010, mouliMECOM2024}, while Rényi and collision entropies tune sensitivity to rare or dominant eigenvalues of $\boldsymbol{\rho}$, supporting tasks such as quantum state tomography \cite{EntropyQuantumTomography2021}.

Across practical quantum networks \ac{QKD}, entanglement-based links, and teleportation, entropic measures set security and rate limits and diagnose noise. Differential entropy addresses uncertainty in continuous-variable (\ac{CV}) systems, whereas Rényi entropies are pivotal to composable security proofs. Accurate entropy models under real noise are needed both to design protocols and to certify performance \cite{Mouli2024Finite_size_QKD_QComm, WangQuantumEntropies2024}. A broad literature compares these measures \cite{gour2021entropy}: von Neumann entropy remains the standard for mixedness and entanglement; Rényi entropies (order $\alpha$) generalize it and appear in hypothesis testing and channel discrimination \cite{wilde2013quantum}; and collision entropy ($\alpha=2$) tightens error bounds and quantifies state overlaps \cite{acharya2020JSAI}. Work on entropy power inequalities and production in open systems, including dynamics under various decoherence models \cite{WangQuantumFidelity2023}, essentially treats \ac{DV} and \ac{CV} systems separately.

It is essential to note that hybrid quantum channels (\ac{HQC}s) \cite{Mouli2025PhDThesis} are highly integrated for practical quantum communication; they are not limited to satellite \ac{QKD}, fiber-network hybrids, or distributed quantum computing. However,  the entropic characterization of such practical use cases remains fragmented due to the lack of a proper mathematical foundation. Existing entropy measures, such as von Neumann or Rényi entropies, are typically derived for idealized noise models, which include pure Gaussian or discrete Pauli channels. Moreover, in practical application scenarios, these models are inadequate in comprehending the interplay of mixed noise sources (e.g., thermal, phase diffusion, and discretization artifacts) \cite{mouli2025SatQDAGMM}. Therefore, it is challenging to determine the performance limits of protocols like \ac{QKD}, quantum error correction, and secrecy estimation in a hybrid noise-aided quantum communication system. Hence, to address this, we statistically model hybrid noise using Gaussian mixture models (\ac{GMM}s), which naturally model discrete-continuous correlations and rigorously analyze differential, Rényi, and collision entropies across this entire landscape. Our research comprehensively demonstrates certain equivalences and insights into how these entropies behave and converge under hybrid conditions, thus carefully laying the groundwork for entropy-based performance and security analysis in next-generation quantum communication systems.

This research work  improves the theoretical and practical insight of entropy in \ac{HQC}s through the following key contributions:

\begin{itemize}
    \item We present a unified entropy framework for \ac{HQC}s incorporating both discrete and continuous variable noise effects. Modeling hybrid noise using \ac{GMM}s captures the joint influence of quantum decoherence and classical distortion in realistic systems such as satellite-based \ac{QKD} and hybrid repeaters.

    \item We provide analytical and numerical studies supported by 3D visualizations of differential, Rényi (\( \alpha = 2 \)), and collision entropies for \ac{HQN}s, revealing their structural behavior and demonstrating the asymptotic equivalence between differential and collision entropy under specific hybrid noise conditions.

    \item We extend the entropy framework to finite-key \ac{QKD}, demonstrating that maintaining entropy estimation errors within a ten percent threshold is crucial for accurate key rate prediction and composable secrecy, thereby establishing the operational relevance of entropy-based analysis for hybrid quantum communication.
\end{itemize}

   Our research advancements lay the groundwork for a theoretical foundation in entropy-based security assessment and resource optimization in hybrid quantum communication networks, with potential applications in \ac{QKD}, quantum sensing, and heterogeneous quantum networking infrastructures.

\textit{Notation,} In this research work, we use $tr(\cdot)$ and $\mathbf{T(\cdot)}$ to denote the trace of a matrix and {Trace-preserving} map, respectively. For a matrix $\mathbf{A}$, $\mathbf{A}^\dag$ and ${\mathbf{A}}^t$ represent the adjoint and transpose, respectively. We denote the complex conjugate of a vector $\boldsymbol{\nu}$ by using $\boldsymbol{\nu}^*$, and the tensor product is denoted by $\otimes$. We represent the Gaussian density by $\mathcal{N}$. For a quantum state $\psi$, ket and bra are denoted by $\ket{\boldsymbol{\psi}}$ and $\bra{\boldsymbol{\psi}}$, respectively. 

\section{System Model}

A qubit is written as $\ket{\boldsymbol{\psi}}=\boldsymbol{\alpha}\ket{\boldsymbol{0}}+\boldsymbol{\beta}\ket{\boldsymbol{1}}$ with $\boldsymbol{\alpha},\boldsymbol{\beta}\in\mathbb{C}$ and $|\boldsymbol{\alpha}|^2+|\boldsymbol{\beta}|^2=1$. Its pure-state density matrix is $\boldsymbol{\rho}=\ket{\boldsymbol{\psi}}\bra{\boldsymbol{\psi}}=\begin{bmatrix}|\boldsymbol{\alpha}|^2 & \boldsymbol{\alpha}^*\boldsymbol{\beta}\\ \boldsymbol{\alpha}\boldsymbol{\beta}^* & |\boldsymbol{\beta}|^2\end{bmatrix}$. Mixed states use $\boldsymbol{\rho}=\sum_i p_i\ket{\boldsymbol{\psi}_i}\bra{\boldsymbol{\psi}_i}$ with $\sum_i p_i=1$ \cite{nielsen_chuang_2010}. The von Neumann entropy $S(\boldsymbol{\rho})=-\mathrm{Tr}(\boldsymbol{\rho}\log\boldsymbol{\rho})$ tracks purity; relative entropy and mutual information quantify distinguishability and total correlations. Noise is modeled by a quantum channel $\mathcal{N}$, a CPTP map with Kraus operators $E_k$: $\mathcal{N}(\boldsymbol{\rho})=\sum_k E_k \boldsymbol{\rho} E_k^\dagger$ \cite{wilde2013quantum}. One can also view it through a system–environment unitary, $\boldsymbol{\rho}\mapsto \operatorname{tr}_E[\,\boldsymbol{U}(\boldsymbol{\rho}\otimes \boldsymbol{\rho}_E)\boldsymbol{U}^\dagger]$. Information carried through the channel is captured by $I(\boldsymbol{\rho};\mathcal{N})$, defined via a supremum of relative entropies over extremal decompositions; a standard communication pipeline applies encoding $\gamma'$, channel action $a'$, and decoding $\delta'$ to map classical inputs to outputs \cite{ohya2004quantum}. For Gaussian channels, the covariance updates as $\boldsymbol{\gamma}\mapsto \boldsymbol{A}^T\boldsymbol{\gamma}\boldsymbol{A}+\boldsymbol{Z}$, where $\boldsymbol{A}$ models amplification, attenuation, or rotation in phase space, and $\boldsymbol{Z}$ adds noise \cite{cerf2007quantum}.

\subsection{Qubit Model}

In quantum mechanics, a pure-state qubit can be represented as a point on the Bloch sphere with coordinates \((\theta, \phi)\), represented by a bivariate function \(\zeta(\theta, \phi)\). Under the presence of noise, small fluctuations act mostly on \(\theta\). However, \(\phi\) remains nearly constant, leading to transformation like \((\theta, \phi) \mapsto (\tilde{\theta}, \phi \pm \delta)\). For mixed states, there is an additional radial coordinate \(r\) contributing to the characterization as \((\theta, \phi, r)\), by a multivariate function \(\tilde{\zeta}(\theta, \phi, r)\). Noise causes small shifts of \(\phi\) and \(r\) with variations \(\delta_1\) and \(\delta_2\) handled as the same for simplicity. For the sake of less complexity, the qubits' motion is approximated by a rotation along a circular or spiral path to facilitate the use of a single scalar variable \(\theta\) for describing the noise-affected behavior in terms of one-dimensional randomness that captures the essential probabilistic characteristics at the reduced computational overhead \cite{mouli2024}.

The hybrid quantum noise \( Z\) is modeled as the sum of  quantum Poissonian noise  \( Z^{(1)} \) and  \ac{AWGN}  \( Z^{(2)} \), such that 
\begin{equation}
    Z= Z^{(1)} + Z^{(2)},
\label{}
\end{equation}
where the  Poissonian noise  \( Z_k^{(1)} \) follows the \ac{p.m.f.} \cite{fox2006quantum}
\begin{equation}
    f_{Z^{(1)}} (j) = \frac{e^{-\lambda}\lambda^j}{j!}, \quad \lambda \geq 0, \quad j \in \{0, 1, 2, \dots\}.
\label{}
\end{equation}
and the  classical AWGN  \( Z^{(2)} \) is Gaussian-distributed with mean \( \mu_{Z^{(2)}} \) and variance \( \sigma_{Z^{(2)}}^2 \).   The convoluted \ac{p.d.f.} represents the hybrid noise \( Z \) affecting the  quantum channel.

\subsection{Quantum Entropies of Hybrid Quantum Noise Model}


The \ac{HQN} can be represented as \cite{mouli2024, Mouli2025ML, Mouli2024Asymp_QKD_SatComm, Mouli2024Finite_size_QKD_QComm, mouliMECOM2024, mouli2025SatQDAGMM},
\begin{equation}
\begin{split}
& f_{Z}(z) = \sum_{i=0}^{R} \frac{e^{-\lambda} \lambda^i}{i!} \frac{1}{\sigma_{Z_2} \sqrt{2\pi}} e^{-\frac{1}{2} \left(\frac{z - i - \mu_{Z_2}}{\sigma_{Z_2}}\right)^2} \\ &= \sum_{i=0}^{R} w_i \, \mathcal{N} \left(z; \mu_i^{(z)}, \, {\sigma_i^{(z)}}^2\right).
\end{split}
\label{eq HQN pdf}
\end{equation}
where \( w_i \) are the mixture weights (such that \( \sum_{i=1}^{R} w_i \approx 1 \)),and  \( \mathcal{N}(z ; \mu_i, \Sigma_i) \) is the Gaussian  \ac{p.d.f.}. The \ac{p.d.f.}  of noise in higher dimensions can be represented using a Gaussian mixture model defined over a random vector  
 $\mathbf{z} \in \mathbb{R}^D$ as
\begin{equation}
f_{\boldsymbol{Z}}(\mathbf{z}) = \sum_{i=0}^{R} w_{i}.\mathcal{N} \Big(\mathbf{z};\boldsymbol{\mu}_{i}^{(z)},\,{\boldsymbol{\Sigma}_{i}^{(z)}}\Big),
\label{eq approx pdf of HQN}   
\end{equation}
where $w_{i}=\frac{e^{-\lambda}\lambda^i}{i!}$ represents the weightage of the Gaussian mixtures, $\mathbf{z}$ is the random vector in $\mathbb{R}^d$, $d$ denotes the dimension of the vector $\mathbf{z}$, $\boldsymbol{\mu}_{i}^{(z)}$ is the mean vector, $\boldsymbol{{\Sigma}}_{i}^{(z)}$ is the covariance matrix of the corresponding Gaussian density $\mathcal{N}$, and the multivariate Gaussian is given by 
\begin{multline}
    \mathcal{N}(\mathbf{z};\boldsymbol{\mu}_{i}^{(z)},\boldsymbol{\Sigma}_{i}^{(z)})
=\\\frac{1}{(2\pi)^{d/2}|\boldsymbol{\Sigma}_{i}^{(z)}|^{1/2}}e^{-\frac{1}{2}(\mathbf{z}-\boldsymbol{\mu}_{i}^{(z)})^{T}\{\boldsymbol{\Sigma}_{i}^{(z)}\}^{-1}(\mathbf{z}-\boldsymbol{\mu}_{i}^{(z)})}.
\label{eq5}
\end{multline}
  With this formulation \eqref{eq5} of \ac{HQN} as a weighted multivariate Gaussian sum, it is now possible to compute information-theoretic measures such as differential entropy. 
The differential entropy of \( \boldsymbol{Z} \), with \ac{p.d.f.} \( f_{\boldsymbol{Z}}(\mathbf{z}) \), over this support \( \chi_{\boldsymbol{Z}} \), is defined as
   \begin{equation}
        H(\boldsymbol{Z}) = \mathbb{E}[-\log f_{\boldsymbol{Z}}(\mathbf{z})] = -\int_{\chi_{\boldsymbol{Z}}} f_{\boldsymbol{Z}}(\mathbf{z}) \log f_{\boldsymbol{Z}}(\mathbf{z}) \, d\mathbf{z}
   \label{eq differential entropy of hybrid quantum noise}
   \end{equation}
This is the most widely used entropic measure in quantum information theory, specifically for quantifying mutual information between a transmitter and receiver in quantum communication systems. Alongside differential entropy, many other meaningful quantum entropy measures include von Neumann entropy, Rényi entropy, and Unified entropy, as discussed within \cite{ohya2004quantum}. For \( \alpha \to 1 \), Rényi entropy converges to differential entropy for \ac{HQN} \cite{imreGyongyosi2012advancedQ_BOOK}. This leads towards the collision entropic measure of \ac{HQN}.

The \( \alpha \)-order quantum Rényi entropy of \ac{HQN} is given by 
   \begin{equation}
       H_\alpha(\boldsymbol{Z}) = \frac{1}{1-\alpha} \log \int f_{\boldsymbol{Z}}(\mathbf{z})^\alpha \, d\mathbf{z}, 
   \label{eq Rényi entropy of hybrid quantum noise}
   \end{equation}
where $0 \leq \alpha < \infty$, $\alpha \neq 1$ , and \( \alpha \) is a non-negative parameter that characterizes the entropy.

The equivalence of collision entropy (also known as Rényi entropy of order 2) and differential entropy for \ac{HQN}'s density function is generally not exact due to fundamental differences in their definitions. However, there are significant relationships between them. The Collision Entropy (Rényi entropy of order 2) of \ac{HQN} is given by  \vspace{-0.25 cm}
   \begin{multline}
       H_2(f_{\boldsymbol{Z}}) = -\log \int f_{\boldsymbol{Z}}(\mathbf{z})^2 d\mathbf{z} \\ = -\log \int \left( \sum_{i=1}^{R} w_i \mathcal{N}(\mathbf{z} ; \boldsymbol{\mu}_i, \boldsymbol{\Sigma}_i) \right)^2 d\mathbf{z}
   \label{collision entropy  HQN}
   \end{multline}
   where \( f_{\boldsymbol{Z}}(\mathbf{z}) \) is the \ac{p.d.f.} of the \ac{HQN} given in \eqref{eq approx pdf of HQN}.
Expanding \( f_{\boldsymbol{Z}}(\mathbf{z})^2 \),  \vspace{-0.25 cm}
\begin{multline}
    f_{\boldsymbol{Z}}(\mathbf{z})^2 = \left( \sum_{i=1}^{R} w_i \mathcal{N}(\mathbf{z} ; \boldsymbol{\mu}_i, \boldsymbol{\Sigma}_i) \right) \left( \sum_{j=1}^{R} w_j \mathcal{N}(\mathbf{z} ; \boldsymbol{\mu}_j, \boldsymbol{\Sigma}_j) \right) \\
    = \sum_{i=1}^{R} \sum_{j=1}^{R} w_i w_j \mathcal{N}(\mathbf{z} ; \boldsymbol{\mu}_i, \boldsymbol{\Sigma}_i) \mathcal{N}(\mathbf{z} ; \boldsymbol{\mu}_j, \boldsymbol{\Sigma}_j).
\end{multline}
Now, we compute the integral  \vspace{-0.25 cm}
\begin{multline}
I = \int f_{\boldsymbol{Z}}(\mathbf{z})^2 d\mathbf{z} \\ = \int \sum_{i=1}^{R} \sum_{j=1}^{R} w_i w_j \mathcal{N}(\mathbf{z} ; \boldsymbol{\mu}_i, \boldsymbol{\Sigma}_i) \mathcal{N}(\mathbf{z} ; \boldsymbol{\mu}_j, \boldsymbol{\Sigma}_j) d\mathbf{z}
\\ = \sum_{i=1}^{R} \sum_{j=1}^{R} w_i w_j \int \mathcal{N}(\mathbf{z} ; \boldsymbol{\mu}_i, \boldsymbol{\Sigma}_i) \mathcal{N}(\mathbf{z} ; \boldsymbol{\mu}_j, \boldsymbol{\Sigma}_j) d\mathbf{z}.
\end{multline}
The integral of two Gaussians gives \cite{ahrendt2005multivariate},
\begin{equation}
\int \mathcal{N}(\mathbf{z} ; \boldsymbol{\mu}_i, \boldsymbol{\Sigma}_i) \mathcal{N}(z ; \boldsymbol{\mu}_j, \boldsymbol{\Sigma}_j) \, d\mathbf{z} = \mathcal{N}(\boldsymbol{\mu}_i ; \boldsymbol{\mu}_j, \boldsymbol{\Sigma}_i + \boldsymbol{\Sigma}_j),
\end{equation}
where the right-hand side is a Gaussian evaluated at \( \boldsymbol{\mu}_i \) under the distribution \( \mathcal{N}(\boldsymbol{\mu}_j, \boldsymbol{\Sigma}_i + \boldsymbol{\Sigma}_j) \), which simplifies to
\begin{equation}
\frac{e^{ \left( -\frac{1}{2} (\boldsymbol{\mu}_i - \boldsymbol{\mu}_j)^T (\boldsymbol{\Sigma}_i + \boldsymbol{\Sigma}_j)^{-1} (\boldsymbol{\mu}_i - \boldsymbol{\mu}_j) \right)}}{(2\pi)^{d/2} |\boldsymbol{\Sigma}_i + \boldsymbol{\Sigma}_j|^{1/2}}.
\end{equation}
Thus, the integral simplifies to
\begin{equation}
I = \sum_{i=1}^{R} \sum_{j=1}^{R} w_i w_j \frac{e^{\left( -\frac{1}{2} (\boldsymbol{\mu}_i - \boldsymbol{\mu}_j)^T (\boldsymbol{\Sigma}_i + \boldsymbol{\Sigma}_j)^{-1} (\boldsymbol{\mu}_i - \boldsymbol{\mu}_j) \right)}}{(2\pi)^{d/2} |\boldsymbol{\Sigma}_i + \boldsymbol{\Sigma}_j|^{1/2}}.
\end{equation}
Taking the logarithm gives  \vspace{-0.25 cm}
\begin{multline}
H_2(\boldsymbol{Z}) \\= -\log \sum_{i=1}^{R} \sum_{j=1}^{R} w_i w_j \frac{e^{\left( -\frac{1}{2} (\boldsymbol{\mu}_i - \boldsymbol{\mu}_j)^T (\boldsymbol{\Sigma}_i + \boldsymbol{\Sigma}_j)^{-1} (\boldsymbol{\mu}_i - \boldsymbol{\mu}_j) \right)}}{(2\pi)^{d/2} |\boldsymbol{\Sigma}_i + \boldsymbol{\Sigma}_j|^{1/2}}.
\label{collision entropy HQN}
\end{multline}
The term \( |\boldsymbol{\Sigma}_i + \boldsymbol{\Sigma}_j|^{1/2} \) controls the spread of the mixture components. In contrast, the term \( e^{(-\frac{1}{2} (\boldsymbol{\mu}_i - \boldsymbol{\mu}_j)^T (\boldsymbol{\Sigma}_i + \boldsymbol{\Sigma}_j)^{-1} (\boldsymbol{\mu}_i - \boldsymbol{\mu}_j))} \) represents the similarity between components if \( \boldsymbol{\mu}_i \) and \( \boldsymbol{\mu}_j \) are close, the exponent is close to 1; if they are far apart, it decays exponentially. The sum over all \( i, j \) accounts for intra- and inter-component interactions.
For well-separated components, that is, when the components are far apart, the dominant terms come from \(i = j\), simplifying the summation structure.


The \ac{HQN} consists of a well-separated \ac{GMM} \cite{mouliMECOM2024}, where the components are far apart relative to their covariances. Under certain conditions, collision and differential entropy expressions become asymptotically equivalent. For well-separated Gaussian components (\(i\neq j \)), the integral is exponentially small, meaning the dominant terms arise from the self-overlap terms (\( i = j \)),  \vspace{-0.25 cm}
\begin{equation}
I \approx \sum_{i=1}^{R} w_i^2 \frac{1}{(2\pi)^{d/2} |\boldsymbol{\Sigma}_i|^{1/2}}.
\end{equation}
Thus, the collision entropy of \ac{HQN} simplifies to  \vspace{-0.25 cm}
\begin{equation}
H_2(\boldsymbol{Z}) \approx -\log \sum_{i=1}^{R} w_i^2 \frac{1}{(2\pi)^{d/2} |\boldsymbol{\Sigma}_i|^{1/2}}.
\label{collision entropy HQN approx}
\end{equation}

\begin{figure*}[!]
    \subfloat[\centering ]{{\includegraphics[width=2.2in]{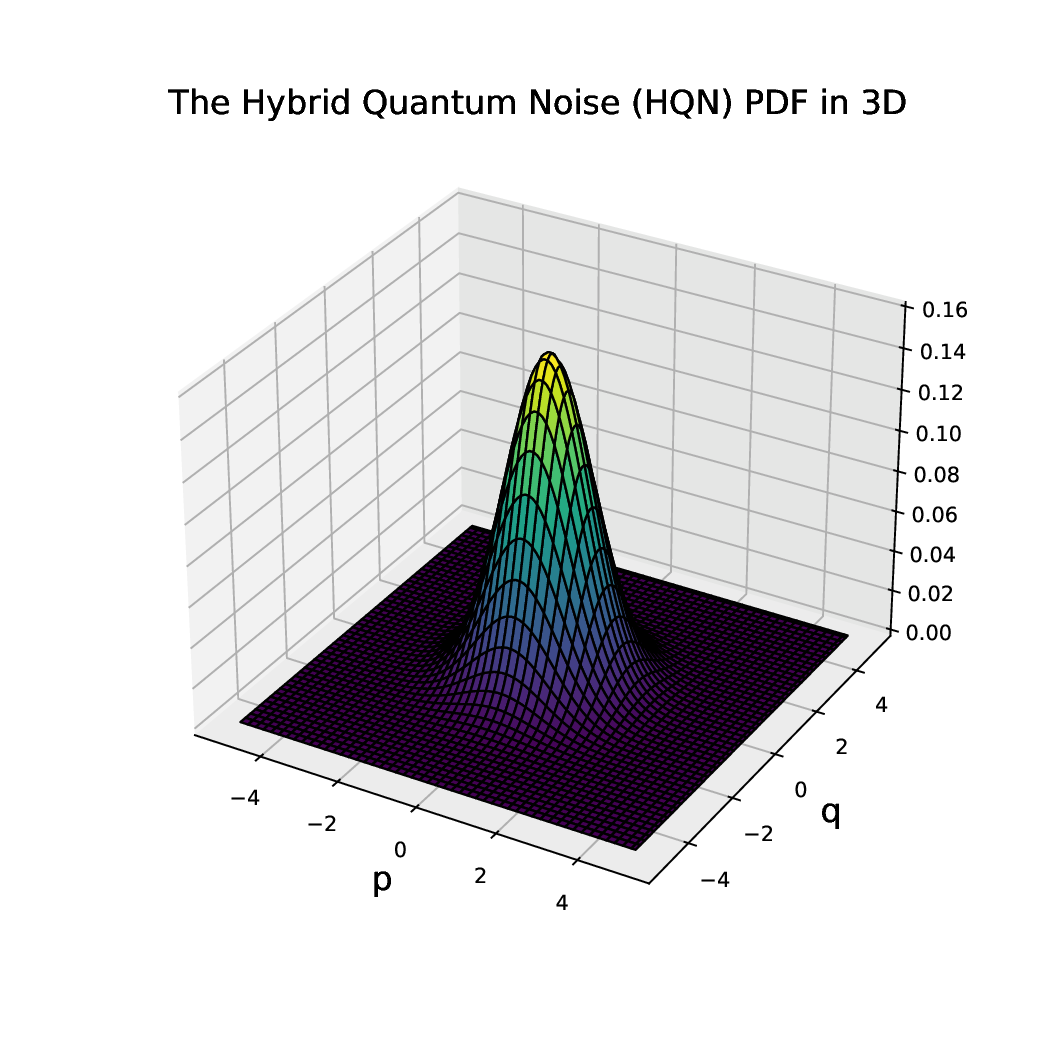} }}%
    \qquad
    \subfloat[\centering ]{{\includegraphics[width=2.2in]{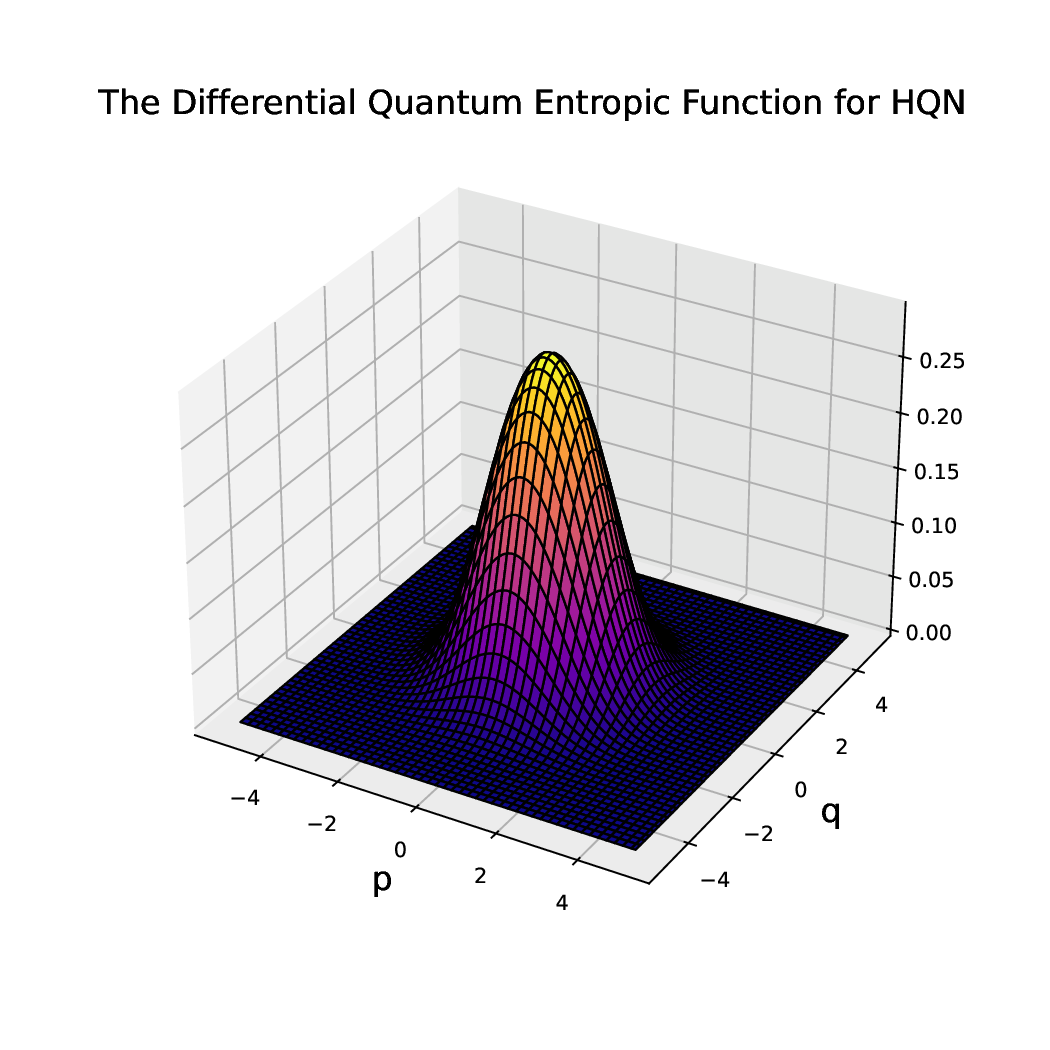} }}
    \qquad
    \subfloat[\centering ]{{\includegraphics[width=2.2in]{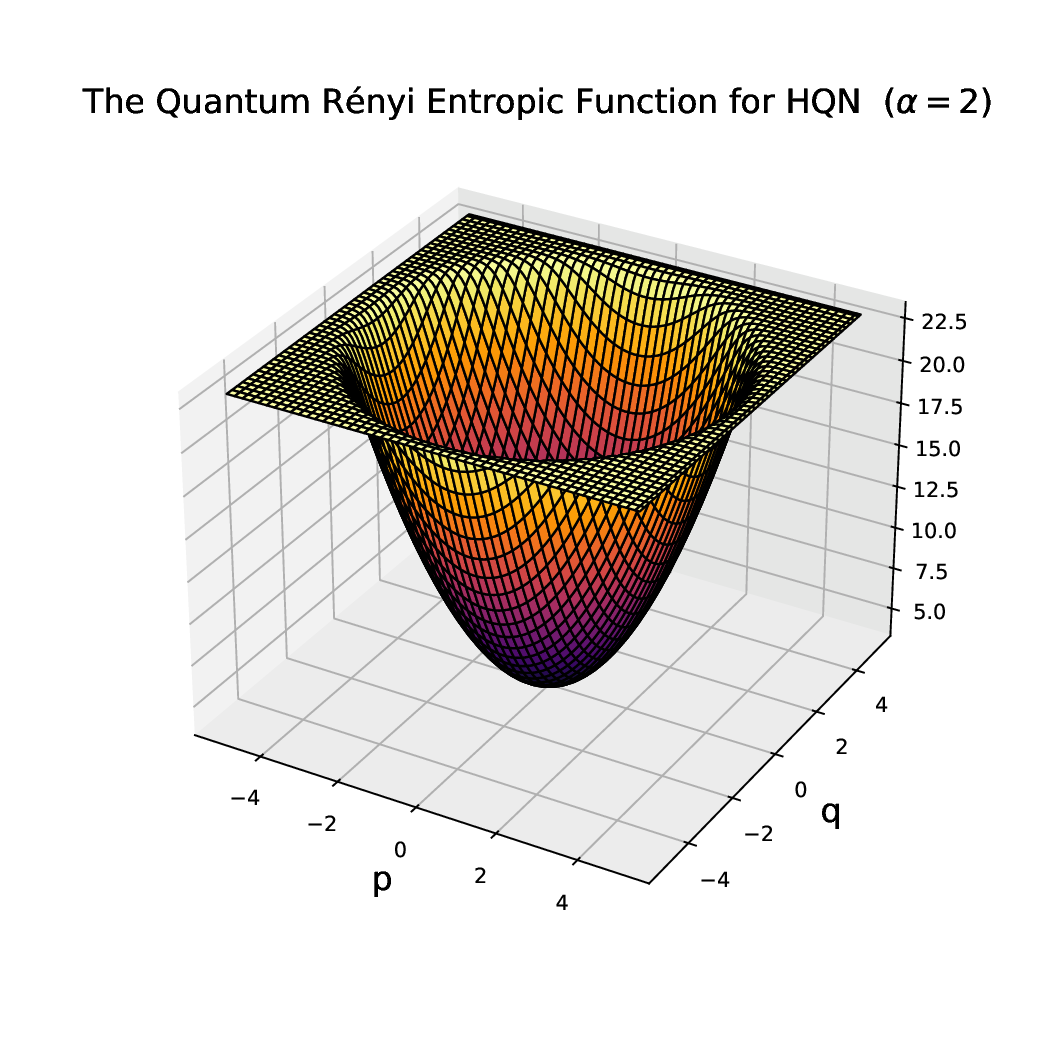} }}
    \caption{(a) 3D representation of Hybrid Quantum Noise modeled using \ac{GMM}, (b) 3D representation of the differential entropic function associated with \ac{HQN}, (c) 3D representation of the quantum Rényi entropic function of HQN.}%
    \label{fig 1}%
    \vspace{-0.5cm}
\end{figure*}
\begin{figure*}[!]
    \subfloat[\centering ]{{\includegraphics[width=2.2in]{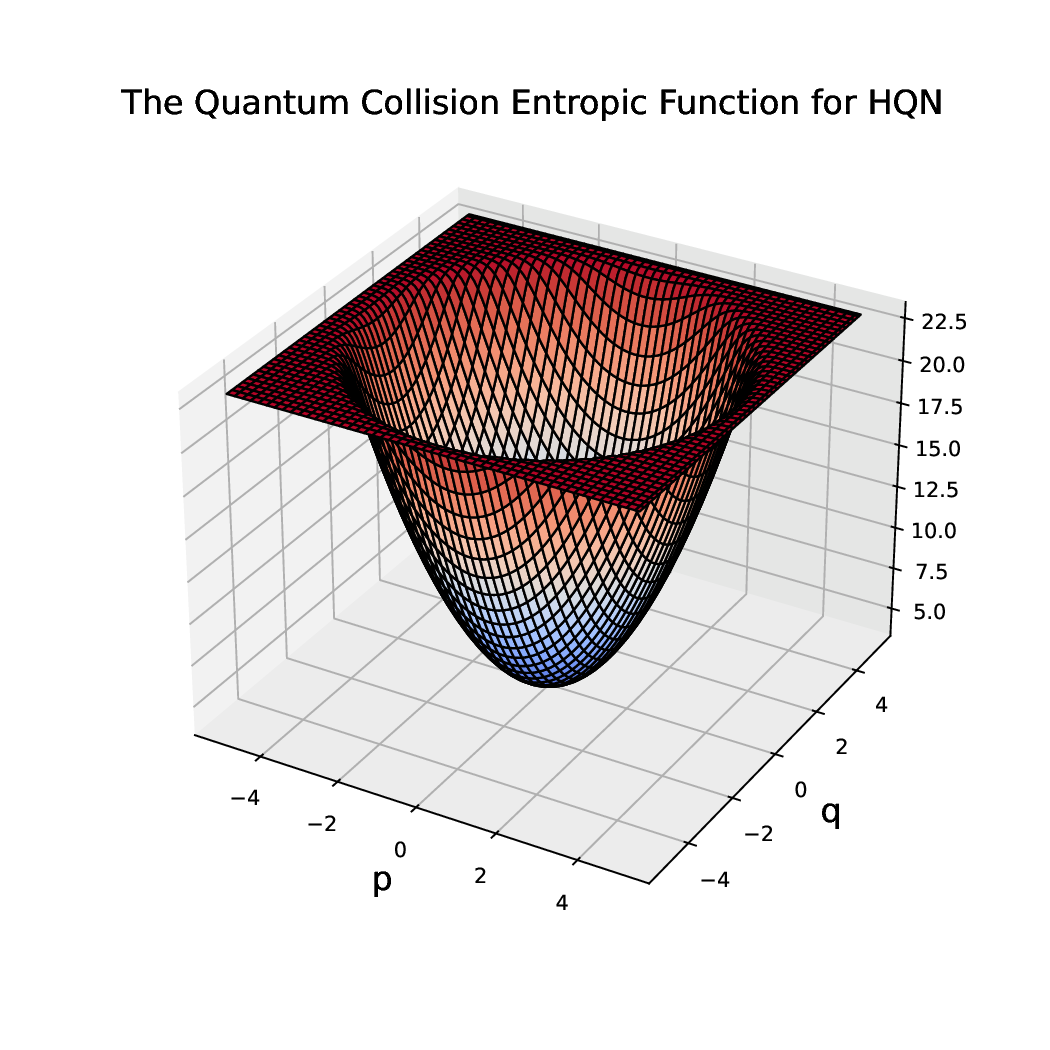} }}%
    \qquad
    \subfloat[\centering ]{{\includegraphics[width=2.2in]{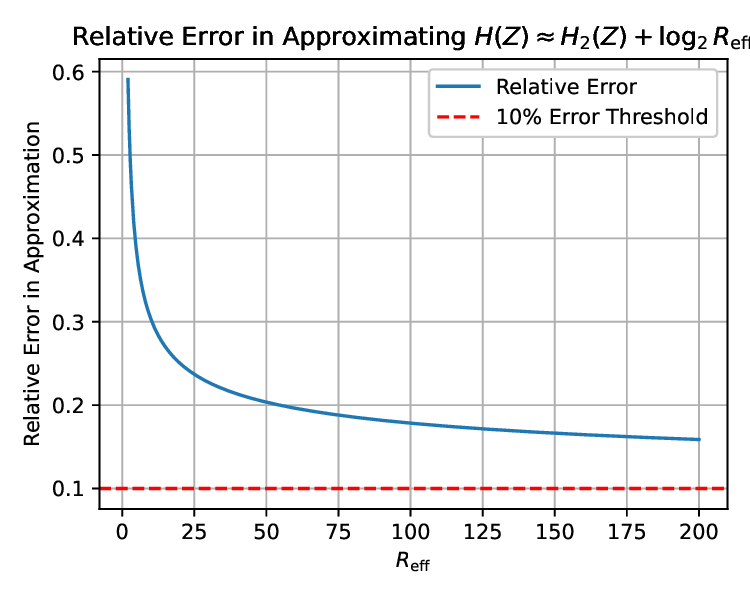} }}
    \qquad
    \subfloat[\centering ]{{\includegraphics[width=2.2in]{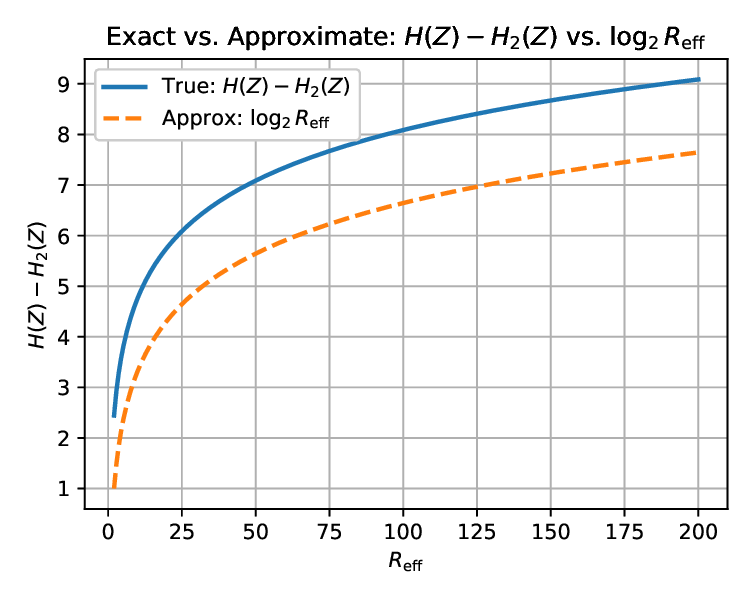} }}
    \caption{(a) 3D representation of the quantum collision entropic function of HQN, (b) Relative error between $H(Z) - H_2(Z) = \log_2 R_{\text{eff}} + \tfrac{d}{2}\log_2 e$ and its approximation $\log_2 R_{\text{eff}}$ versus effective rank $R_{\text{eff}}$ for $d=2$. The red dashed line marks the 10\% error threshold where the approximation is accurate, (c) Comparison of $H(Z) - H_2(Z)$ with its approximation $\log_2 R_{\text{eff}}$ for varying $R_{\text{eff}}$ at $d=2$. The approximation approaches the exact value as $R_{\text{eff}}$ increases, confirming its validity in high-rank regimes. }%
    \label{fig 2}%
\end{figure*}

\subsection{Asymptotic Equivalence of Collision Entropy and Differential Entropy for \ac{HQN}}

For well-separated components $\mathcal{N}$, the dominant contribution to \( f_{\boldsymbol{Z}}(\mathbf{z}) \log f_{\boldsymbol{Z}}(\mathbf{z}) \) comes from within each Gaussian component  \vspace{-0.25 cm}
\begin{equation}
H(\boldsymbol{Z}) \approx \sum_{i=1}^{R} w_i H(\mathcal{N}_i) - \sum_{i=1}^{R} w_i \log w_i.
\end{equation}
where $H(\cdot)$ is the differential entropy function.
The detailed explanation of this approximation is shown in the Appendix.
Since for a Gaussian $H(\mathcal{N}_i) = \frac{d}{2} \log (2\pi e) + \frac{1}{2} \log |\boldsymbol{\Sigma}_i|$,\cite{jwo2023geometric}
we obtain
\begin{multline} 
H(\boldsymbol{Z}) \approx \sum_{i=1}^{R} w_i \left( \frac{d}{2} \log (2\pi e) + \frac{1}{2} \log |\boldsymbol{\Sigma}_i| \right) - \sum_{i=1}^{R} w_i \log w_i
\\ \approx \frac{d}{2} \log (2\pi e)
+ H(\mathbf{w}),
 \end{multline}
where \( H(\mathbf{w}) = -\sum_{i=1}^{R} w_i \log w_i \) is the discrete Shannon entropy of the mixture weights, $\sum_{i=1}^{R} w_i =1$ and $\Sigma_{i}=\mathbf{I}_{d}$.

Since \( \sum w_i^2 \) is related to the effective number of components (denoted as \( R_{\text{eff}} \)) \cite{mouliMECOM2024, Mouli2025ML}, we approximate $\sum_{i=1}^{R} w_i^2 \approx \frac{1}{R_{\text{eff}}}$. Thus
\begin{equation}
H_2(\boldsymbol{Z}) \approx 
\frac{d}{2} \log (2\pi). 
\label{collision entropy final}
\end{equation}
A detailed overview of this approximation is given in the Appendix.
Similarly,
\begin{multline}
    H(\boldsymbol{Z}) \approx \frac{d}{2} \log (2\pi e) + 
H(\mathbf{w})
\approx \frac{d}{2} \log (2\pi e) 
+ \log R_{\text{eff}},
\label{differential entropy final}
\end{multline}
using the approximation \( H(\mathbf{w}) \approx \log R_{\text{eff}} \), as $ \sum_i w_i \log w_i \leq \log \left( \sum_i w_i^2 \right)=-\log R_{\text{eff}}$. Now comparing \eqref{collision entropy final} with \eqref{differential entropy final},
the difference \vspace{-0.25 cm}
\begin{equation}
H(\boldsymbol{Z}) - H_2(\boldsymbol{Z}) \approx  \log_2 R_{\text{eff}} + \frac{d}{2} \log_2 e.
\end{equation}
For large \( R_{\text{eff}} \), this shows that  \vspace{-0.1 cm}
\begin{equation}
H(\boldsymbol{Z}) \approx H_2(\boldsymbol{Z}) + \log_2 R_{\text{eff}}.
\end{equation}


Hence, differential entropy is typically preferred over collision entropy in quantum communication due to its sensitivity to system changes and disturbances within the system. Differential entropy yields higher values than collision entropy, measuring a broader range of statistical uncertainty. Additionally, it considers the maximized disturbance to the quantum system and a superior selection of measures when accounting for information loss, noise, or decoherence in continuous-variable quantum systems. 

\vspace{-0.1 cm}
\section{Application in QKD Estimation}
\vspace{-0.1 cm}
\textcolor{black}{A finite-size Gaussian-modulated CV-QKD protocol with reverse reconciliation is considered, showing how the collision entropy enters the finite-key bound \cite{Mouli2024Finite_size_QKD_QComm}; the $10\%$ threshold is used as a conservative illustrative stress-test due to the exponential sensitivity of finite-key security to entropy errors, while $R$ is chosen to resolve the effective noise rank (ensuring entropy saturation) and the Poisson parameter governs noise complexity and entropy variation \cite{Mouli2025ML}.
}
Therefore, a $10\%$ approximation error in the conditional collision entropy $H_2(X|E)$ has significant implications for finite-regime QKD, as it directly affects both the estimated Eve’s success probability and the secure key length. In finite-key security analysis, Eve’s optimal success probability satisfies
\begin{equation}
p_{\mathrm{succ}}(\mathrm{Eve}) \le 2^{-H_2(X|E)}.
\end{equation}
If the estimated entropy $\widehat{H}_2 = (1+\delta)H_2$ deviates by a relative error $\delta$, the ratio between the estimated and true bounds becomes  \vspace{-0.35 cm}
\begin{equation}
\frac{p_{\mathrm{succ}}(\widehat{H}_2)}{p_{\mathrm{succ}}(H_2)} = 2^{-\Delta}, \quad \text{where} \quad \Delta = \delta H_2.
\end{equation}
Hence, a $10\%$ underestimation ($\delta = -0.1$) with $H_2=100$ bits yields a degradation factor of $2^{10}\!\approx\!10^3$, loosening Eve’s bound by three orders of magnitude. For key extraction, the Leftover Hash Lemma bounds the secret key length as
\begin{equation}
\ell \le H_2^{\varepsilon_s}(X|E) - \mathrm{leak}_{\mathrm{EC}} - 2\log_2\!\left(\frac{1}{2\varepsilon_{\mathrm{PA}}}\right),
\end{equation}
where $\varepsilon_s$ and $\varepsilon_{\mathrm{PA}}$ are smoothing and privacy amplification parameters. Denoting the total penalty by $\Pi$, the true margin is $m = H_2 - \Pi$. Incorporating entropy error gives
\begin{equation}
\ell_{\text{est}} - \ell_{\text{true}} = \Delta = \delta H_2,
\end{equation}
and the relative deviation becomes
\begin{equation}
\frac{\ell_{\text{est}} - \ell_{\text{true}}}{\ell_{\text{true}}} = \frac{\delta H_2}{H_2 - \Pi}.
\end{equation}
When $m$ is small, as in finite-size regimes, even a $10\%$ error can nullify the key ($\ell_{\text{est}}\le0$) or falsely enhance security. Consequently, maintaining the approximation error of $H_2(X|E)$ within $10\%$ ensures composable secrecy and accurate finite-key rate estimation.

\begin{figure}[t]
\centering
\includegraphics[width=1\columnwidth]{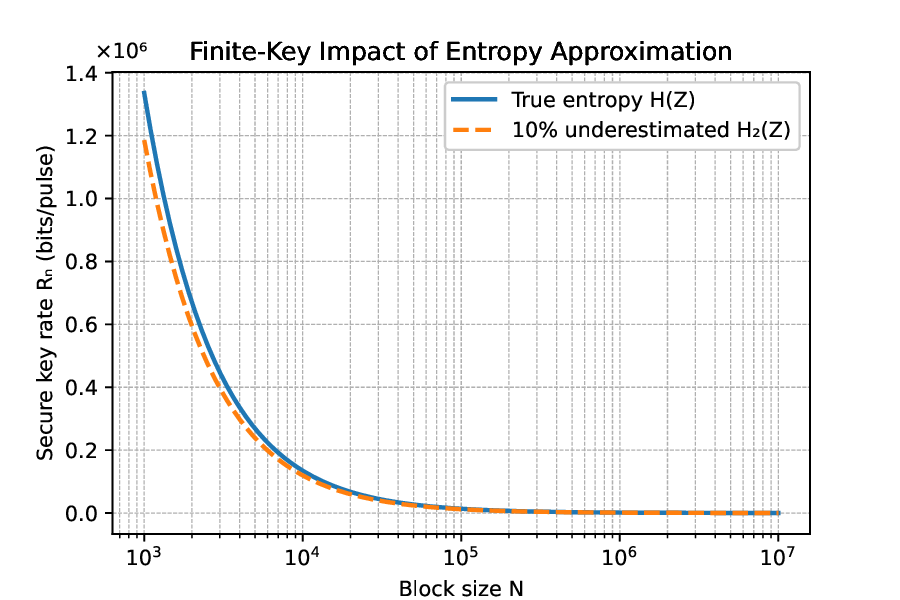}\vspace{-0.0cm}
\caption{Finite-regime entropy approximation for secure key rate $R_{N}$ under different block size $N$ for true $H(Z)$ and 10\% underestimated $H_{2}(Z)$.} 
\vspace{-0.5cm}
\label{Fig. 3}
\end{figure}

\section{Numerical Analysis}
In numerical analysis, the quantum entropic function is evaluated over the quadrature space, where the $p$ and $q$ axes represent phase-space coordinates of the quantum state. The conjugate quadratures $p$ and $q$ describe the quantum state of a harmonic oscillator in the \ac{CV} regime and are used to characterize quantum optical modes such as coherent, squeezed, and Gaussian states for information encoding. In quantum optics, the annihilation and creation operators relate to the position and momentum operators as
$\hat{q} = \frac{1}{\sqrt{2}}(\hat{a} + \hat{a}^\dagger), \quad
\hat{p} = \frac{1}{i\sqrt{2}}(\hat{a} - \hat{a}^\dagger),$ satisfying the canonical commutation relation $[\hat{q}, \hat{p}] = i$. Hence, a quantum state can be represented by its density distribution over the $(q,p)$ plane. In \ac{CV} quantum computation, information is encoded in quadrature amplitudes rather than discrete basis states~\cite{HanzoSatcomm2019}. The 3D plot thus visualizes entropy variations across phase-space configurations, revealing the structure and spread of quantum uncertainty.

We numerically analyze this \ac{HQN} model, visualizing its probabilistic and entropic structure using \ac{GMM}s over the $p,q$ quadrature space. In Fig. \ref{fig 1} (a), the 3D representation catches the \ac{HQN} density distribution, combining discrete Poissonian noise with continuous-variable Gaussian noise, a characteristic of hybrid quantum communication systems. These quadratures relate the position and momentum operators within quantum continuous-variable regimes to create an entropy-evaluation phase-space basis. We utilize various quantum entropic functions on the density distribution to quantify uncertainty. Fig. \ref{fig 1} (b) presents a differential entropic function of this kind, and its surface depicts the pointwise entropy contributions all across that $p,q$ domain. It yields its integral from the entire total differential entropy, thereby reflecting the complete uncertainty within the system given hybrid noise. Fig.\ref{fig 1} (c) illustrates a 3D representation of the quantum Rényi entropic function of \ac {HQN} for the generalization of the uncertainty of the quantum channel. This portrayal illustrates how the entropic landscape evolves under second-order conditions for Rényi entropy, providing a broad perspective on the information content and distinguishability within quantum states subjected to hybrid noise.

Fig.~\ref{fig 2}(a) shows the collision entropic function of \ac{HQN}, illustrating the degree of state overlap and indistinguishability. Using the HQN model based on GMM, the measure captures the impact of hybrid noise on the robustness and separability of quantum states within phase space. Beyond asymptotic conditions, collision entropy plays a crucial role in finite-key \ac{QKD}, where statistical fluctuations from limited data must be considered. It provides a more conservative and experimentally verifiable bound than Shannon entropy, which represents a typical-case scenario. Fig.~\ref{fig 2}(b) presents the relative error between the exact and approximate difference of differential and collision entropies, $H(Z)-H_{2}(Z)$ and $\log_{2}R_{\text{eff}}$, respectively, as a function of the effective rank $R_{\text{eff}}$. The red dashed line marks the 10\% relative error threshold that defines the approximation accuracy. This validates the theoretical relation derived in (21), showing that $H(Z)\!\approx\!H_{2}(Z)+\log_{2}R_{\text{eff}}$. As $R_{\text{eff}}$ increases, the additive correction term $\frac{d}{2}\log_{2}e$ becomes negligible, and the approximation $H(Z)\!\approx\!\log_{2}R_{\text{eff}}$ becomes tighter. Fig.~\ref{fig 2}(c) compares $H(Z)-H_{2}(Z)$ with $\log_{2}R_{\text{eff}}$ for $d=2$, demonstrating convergence of the empirical results to the theoretical approximation as $R_{\text{eff}}$ grows. These results confirm the validity of the entropy estimation model $H(Z)-H_{2}(Z)\!\approx\!\log_{2}R_{\text{eff}}$ and quantify the error bounds and convergence behavior of entropy approximations in hybrid quantum systems.

Fig. \ref{Fig. 3} extends the analysis by quantifying how approximation inaccuracies in $H(Z)\!\approx\! H_{2}(Z)+\log_{2}R_{\mathrm{eff}}$ influence finite-key QKD performance. In Fig. \ref{Fig. 3}, the secure key rate $R_{N}$ is plotted against the block size $N$ for both the exact differential entropy and its 10\% underestimated collision-entropy approximation. The results clearly show that even small entropy deviations cause significant degradation in $R_{N}$ for short block sizes, emphasizing the importance of maintaining precise entropy estimation in finite-sample regimes. As $N$ increases, the two curves converge, validating the asymptotic equivalence derived in (22).

This also impacts computational efficiency in quantum communication protocols, including \ac{QKD} and quantum sensing, for which entropy estimation in resource-constrained settings, and in real-time, is critical. These tested assumptions yield straightforward safety gauges, keeping strong theories, and measured applications within detailed quantum systems. 

\vspace{-0.10 cm}

\section{Conclusion}

This work investigated the relationship between the quantum Rényi entropy of order 2 and differential entropy within the framework of effective rank and dimensionality. Through an analysis of the asymptotic behavior of their difference \( H(Z) - H_2(Z) \), we demonstrated that the approximation \( H(Z) \approx H_2(Z) + \log_2 R_{\text{eff}} \) becomes increasingly accurate as the effective rank grows, with the additive correction term \( \frac{d}{2} \log_2 e \) becoming negligible in size. This finding provides a valuable guide for estimating entropy in mixed, high-dimensional quantum states, supporting the use of simpler entropic forms in large-scale quantum systems. The QKD analysis further revealed that maintaining entropy estimation errors within a ten percent threshold is critical for preserving accurate secure key rates and composable secrecy in finite-size regimes. Overall, our analysis links operational entropy measures with structural features in quantum states, offering a scalable perspective for applications in quantum information processing, learning-based quantum models, and thermodynamic inference. \vspace{-0.25 cm}

\section{Appendix}
\subsection{{Approx. Differential Entropy of \ac{HQN}}}
When the Gaussian components are well-separated, their overlaps are negligible. So, at most points \(\mathbf{z} \), only one \( \mathcal{N}_i(\mathbf{z}) \) dominates \( f_{\boldsymbol{Z}}(\mathbf{z}) \). This motivates the approximation,\vspace{-0.1 cm}
\begin{equation}
f_{\boldsymbol{Z}}(\mathbf{z}) \log f_{\boldsymbol{Z}}(\mathbf{z}) \approx \sum_{i=1}^R w_i \mathcal{N}_i(\mathbf{z}) \log (w_i \mathcal{N}_i(\mathbf{z}))
\end{equation}
where $\mathcal{N}_i(\mathbf{z})=\mathcal{N}(\mathbf{z} ; \boldsymbol{\mu}_i, \boldsymbol{\Sigma}_i)$ is the $i$th Gaussian component. Then the differential entropy becomes approximately, \vspace{-0.1 cm}
\begin{equation}
\begin{aligned}
H(\boldsymbol{Z}) &= - \int f_{\boldsymbol{Z}}(\mathbf{z}) \log f_{\boldsymbol{Z}}(\mathbf{z}) \, d\mathbf{z} \\
&\approx - \sum_{i=1}^R w_i \int \mathcal{N}_i(\mathbf{z}) \log (w_i \mathcal{N}_i(\mathbf{z})) \, d\mathbf{z} \\
&= - \sum_{i=1}^R w_i \left[ \log w_i \int \mathcal{N}_i(\mathbf{z}) d\mathbf{z} + \int \mathcal{N}_i(z) \log \mathcal{N}_i(\mathbf{z}) d\mathbf{z} \right] \\
&= - \sum_{i=1}^R w_i \log w_i + \sum_{i=1}^R w_i H(\mathcal{N}_i)
\end{aligned}
\end{equation}
Hence,
$H(\boldsymbol{Z}) \approx \sum_{i=1}^{R} w_i H(\mathcal{N}_i) - \sum_{i=1}^{R} w_i \log w_i$
The term \( \sum w_i H(\mathcal{N}_i) \) is the weighted average of component entropies, and \( -\sum w_i \log w_i \) is the entropy of the discrete weight distribution (Poisson). These ensure an well approximation for \( \| \mu_i - \mu_j \| \gg \sqrt{\text{tr}(\boldsymbol{\Sigma}_i + \boldsymbol{\Sigma}_j)} \) for all \( i \ne j \), and in low-overlap \ac{GMM}s.

\subsection{Approx. Collision Entropy of \ac{HQN} }
The \, collision  \, entropy of \ac{HQN} \vspace{-0.2 cm}
\begin{multline}
    H_2(f) \approx -\log \sum_{i=1}^{R} w_i^2  \frac{1}{(2\pi)^{d/2} |\Sigma_i|^{1/2}}\\
    = \frac{d}{2} \log (2\pi) -\log \sum_{i=1}^{R} w_i^2 |\Sigma_i|^{-1/2}.
\end{multline}
Since \( \log \) is concave, and utilizing Jensen's inequality,  \vspace{-0.2 cm}
\begin{multline}
\log \left( \sum_i w_i^2  |\Sigma_i|^{-1/2} \right)= \log  \left( 
\mathbb{E}_{{w}^2}[|\Sigma_i|^{-1/2}]\right) \\ \leq  
\mathbb{E}_{{w}^2}[\log \left(|\Sigma_i|^{-1/2}\right)] =
 \sum_i w_i^2  \log \left(  |\Sigma_i|^{-1/2} \right).
\end{multline}
This is exact when all \( |\Sigma_i| \) are equal, and in our case $ \Sigma_i = \mathbf{I}_d$ is the identity matrix of order $d$ \cite{mouli2024, mouliMECOM2024, Mouli2025ML, Mouli2024Asymp_QKD_SatComm, Mouli2024Finite_size_QKD_QComm}. 
So,
\begin{equation}
     H_2(f) 
\approx \frac{d}{2} \log(2\pi) 
\end{equation}

\begin{acronym}
\acro{HQC}{hybrid quantum channel}
\acro{HQNM}{hybrid quantum noise model}
\acro{HQN}{hybrid quantum noise}
     \acro{RSA}{Rivest, Shamir and Adelman algorithm}
     \acro{ECC}{elliptic curve cryptography}
     \acro{LOQC}{linear optical quantum computing}   
     \acro{CPTP}{completely positive trace-preserving}
     \acro{MAP}{maximum a posteriori}
    \acro{ML}{machine learning}
    \acro{DL}{deep learning}
    \acro{RL}{reinforcement learning}
    \acro{SVM}{support vector machines}
    \acro{PCA}{principal component cnalysis}
    \acro{KF}{Kalman Filter}
    \acro{DBSCAN}{density-based spatial clustering of applications with noise}
    \acro{WCSS}{within-cluster sum of squares}
    \acro{QML}{quantum machine learning}
    \acro{NN}{neural network}
    \acro{EM}{expectation-maximization}
    \acro{GM}{Gaussian mixture}
    \acro{p.d.f.}{probability density function}
    \acro{p.m.f.}{probability mass function}
    \acro{SNR}{signal-to-noise ratio}
    \acro{SPS}{single-source photon}
    \acro{GMM}{Gaussian mixture model}
    \acro{GMs}{Gaussian mixtures}
    \acro{BIC}{Bayesian information criterion }
    \acro{AIC}{Akaike information criterion}   
    \acro{AWGN}{additive-white-Gaussian noise}
    \acro{OGS}{optical ground station}
    \acro{SKR}{secret key rate}
    \acro{QKD}{quantum key distribution}
    \acro{PNS}{photon number splitting}
    \acro{CV-QKD}{continuous-cariable quantum key distribution}
    \acro{FSO}{free-space optics}
    \acro{MDI}{measure-device-independent}
    \acro{DV-QKD}{discrete-variable quantum key distribution}
    \acro{CV}{continuous variables}
    \acro{DV}{discrete variables}
    \acro{DR}{direct-reconciliation}
    \acro{RR}{reverse-reconciliation}
    \acro{CPTP}{completely positive, trace preserving}
    \acro{PM}{prepare-and-measure}
    \acro{PDTC}{probability distribution of transmission coefficient}
    \acro{PDTE}{probability distribution of transmission efficiency}
    \acro{S/C}{spacecraft}
    \acro{GS}{ground station}
    \acro{LDPC}{low-density parity check}

\end{acronym}

\bibliographystyle{IEEEtran}
\bibliography{IEEEabrv,paper}

\end{document}